\documentstyle[agupp]{article}
\def\pd{\partial}
\def\stru{\displaystyle\rule[-.8ex]{0ex}{2.9ex}}
\lefthead{RUFFOLO ET AL.}
\righthead{DECONVOLUTION OF INTERPLANETARY TRANSPORT}
\received{January~28,~1998}
\revised{March~12,~1998}
\accepted{April~15,~1998}
\paperid{98JA01290}
\cpright{AGU}{1998}
\ccc{0148-0227/98/98JA-01290\$09.00}

\authoraddr{D. Ruffolo and W. Youngdee, Department of Phys\-ics, 
Chulalongkorn University, Bang\-kok 10330, Thailand.
(e-mail: david@astro.phys.sci.chula.ac.th)}

\authoraddr{T. Khumlumlert, Department of Physics, Naresuan
University, Phitsanulok 65000, Thailand.
(e-mail: thiraneek@nu.ac.th)}

\slugcomment{To appear in the {\it Journal 
of Geophysical Research}, 1998.}

\makeatletter
\let\reset@font\empty
\makeatother

\begin{document}

\title{Deconvolution of interplanetary transport of solar energetic
particles}
\author{D. Ruffolo, T. Khumlumlert,\altaffilmark{1} and W. Youngdee}
\affil{\mbox{\ } \\ Department of Physics, Chulalongkorn University,
Bangkok, Thailand}
\altaffiltext{1}{Now at Department of Physics, Naresuan University,
Phitsanulok, Thailand.}

\begin{abstract}
We address the problem of deconvolving the effects of
interplanetary transport on observed intensity and anisotropy profiles 
of solar energetic particles with the goal of determining the time
profile and spectrum of particle injection near the Sun as well as
the interplanetary scattering mean free path.  Semi-au\-to\-ma\-ted
techniques have been developed to quantitatively determine 
the best fit injection profile, assuming (1) a general piecewise linear
profile or (2) a Reid profile of the form 
$[C/(t-t_0)]\exp[-A/(t-t_0)-(t-t_0)/B]$.  The two assumptions for the 
form of the injection profile yielded similar results when we
tested the techniques using ISEE 3 proton data from the
solar flare events of July 20, 1981 (gradual flare), and January 2, 1982 
(impulsive flare).  For the former event, the duration of injection was
much shorter for protons of higher energy (75-147 MeV), which may 
be interpreted as indicating that the coronal mass ejection-driven shock
lost the ability to accelerate protons to $\sim$100 MeV after
traveling beyond a certain distance from the Sun.
\end{abstract}

\begin{article}
\section{1. Introduction}

One of the basic questions one can ask about energetic particles
produced by solar activity is how many are accelerated as a function of
energy and time.  Electromagnetic and neutron diagnostics of interacting 
energetic particles can provide a wealth of information on the particle 
spectrum and the timing of particle acceleration at the flare site 
[e.g., \markcite{{\it Chupp,} 1990;} 
\markcite{{\it Aschwanden et al.,} 1994;} \markcite{{\it Aschwanden,} 1996;} 
\markcite{{\it Debrunner et al.,} 1997;} 
\markcite{{\it Rank et al.,} 1997a, b}].
Complementary information can be obtained from spacecraft or
ground-based observations of escaping, energetic, charged particles.
The interpretation of these direct observations is complicated by the
effects of interplanetary transport, particularly 
scattering due to irregularities in the interplanetary magnetic field
[\markcite{{\it Meyer et al.,} 1956}].  To date, such observations have
always been performed at distances approximately 0.3 AU or greater 
from the Sun, while the mean-free path of interplanetary 
scattering parallel to the magnetic field, $\lambda_\parallel$,
is typically between 0.08 and 0.3 AU (near the Earth),
though it can be $>1$AU for infrequent ``scatter-free'' events
[\markcite{{\it Palmer,} 1982}].
Since the distance from the Sun to the observer is of the same order
of magnitude as, and usually larger than, the mean-free path, 
the interplanetary transport is largely diffusive in most cases,
making it difficult to determine the underlying time profile of
injection near the Sun and possibly affecting the measurement of the
spectrum.

Why should we go through the trouble of deconvolving the effects
of interplanetary transport to determine the injection profile of
escaping energetic particles, when electromagnetic or neutron
diagnostics of interacting particles can yield such information
more directly? A key motivation is that it has recently become widely
accepted that for many flare events the interacting and escaping
energetic particles are accelerated in different locations and
possibly by different mechanisms.
Twenty years ago, it was recognized from X ray observations that
solar flares can be basically classified into two groups
[\markcite{{\it Pallavicini et al.,} 1977}], which are
now commonly called ``impulsive'' and ``gradual'' according to the
duration of their X ray emission.  More recently, 
flares in these two categories have been found to exhibit many
other differences in their physical properties and particle
emissions, although measurements of ionic charge states indicated that
ions from both types of flares are accelerated out of coronal
material [\markcite{{\it Luhn et al.,} 1984, 1987}].
One key difference is that gradual flares are much more
likely to be associated with coronal mass ejections (CMEs),
which in turn often drive a traveling interplanetary shock 
[\markcite{{\it Sheeley et al.,} 1983}].  
For over a decade there has been mounting circumstantial evidence that for 
flare/CME events, the flare is responsible for accelerating
interacting particles, while the CME and/or associated shock is
responsible for accelerating escaping ions on open magnetic field
lines over a wide range of heliolongitudes [e.g., 
\markcite{{\it Mason et al.,} 1984;}
\markcite{{\it Lee and Ryan,} 1986;} \markcite{{\it Reames,} 1990}]. 
(The origin of escaping electrons is subject to debate, but there is
some evidence that they come from the flare site [\markcite{{\it Dr\"oge
et al.,} 1990a}].)  Further evidence against the alternative and 
previously accepted view that escaping ions are accelerated
at the flare site followed by transport within the corona is provided
by recent Solar, Anomalous, and Magnetospheric Particle Explorer 
(SAMPEX) measurements of the charge states of escaping ions
[\markcite{{\it Leske et al.,} 1995;} \markcite{{\it Mason et al.,}
1995;} \markcite{{\it Oetliker et al.,} 1997}], which rule
out transport within the corona for longer than $\sim$1
min [\markcite{{\it Ruffolo,} 1997}].
Therefore observations of escaping ions can provide unique
information about the acceleration of ions out of coronal material 
in the CME/shock region [\markcite{{\it Kahler et al.,} 1990}].

For impulsive flare events, the remarkable enhancement of the
$^3$He/$^4$He ratio among escaping ions [\markcite{{\it Hsieh and
Simpson,} 1970}] by up to a factor of thousands is best explained by
turbulent acceleration in the flare region [\markcite{{\it Fisk,}
1978}].  The observed longitudinal distribution of
solar energetic particles from impulsive events (albeit narrower
than for gradual events) could be explained by interplanetary 
transport perpendicular to the average magnetic field or by transport within
the corona.  Thus, for impulsive events,
a deconvolution of the effects of interplanetary
transport along the field could be useful for characterizing the
process of azimuthal transport, especially if one studies data from
multiple spacecraft at different locations.

Previous analyses of the intensity and/or anisotropy of solar
energetic particles versus energy and time 
for various solar flare/CME events have yielded information on (1) the
injection of particles near the Sun versus energy and time and (2) the
mean-free path of interplanetary scattering.  In most cases, both must
be determined to provide a good fit to the data, although some
previous studies have reported adequate fits by assuming an
instantaneous injection of particles for some events.  

Previous fits to solar energetic particle observations 
have employed various assumptions about interplanetary
transport.  Some authors have restricted their analyses to
cases where the mean-free path is believed to be long and have
neglected interplanetary transport effects altogether 
[e.g., \markcite{{\it Debrunner et al.,} 1988;}
\markcite{{\it Kahler et al.,} 1990;} \markcite{{\it Kallenrode
and Wibberenz,} 1991}].  Analytic solutions employing the diffusive  
approximation [\markcite{{\it Parker,} 1963}] were used by
\markcite{{\it Wibberenz et al.} [1989]}, 
and approximate analytic solutions of an equation of focused transport
[\markcite{{\it Earl,} 1976}] were used for a comprehensive survey of
solar energetic particle observations by \markcite{{\it Ma Sung and
Earl} [1978]}.  \markcite{{\it Lockwood et al.}\ [1982]} used a Monte 
Carlo method to solve the hard-sphere pitch angle scattering model of 
\markcite{{\it Fisk and Axford} [1969]}.
Numerical solutions of a diffusive model that incorporates solar wind
convection and adiabatic deceleration [\markcite{{\it Hamilton,}
1977}] were also employed by \markcite{{\it Beeck and Wibberenz}
[1986]} and \markcite{{\it Beeck et al.}\ [1987]}.
Numerical solutions of the equation of focused transport of 
\markcite{{\it Earl} [1976]} by an eigenfunction expansion technique
were used by \markcite{{\it Bieber et al.}\ [1980, 1986]}, and
solutions using a finite difference method
[\markcite{{\it Ng and Wong,} 1979;} \markcite{{\it Schl\"uter,}
1985}] have also been employed [e.g., \markcite{{\it Dr\"oge et al.,}
1990b;} \markcite{{\it Kallenrode et al.,} 1992}].

Naturally, a deconvolution that employs a less accurate transport 
model should yield a less accurate injection function.  In
particular, our results indicate that
deconvolution with a less accurate transport model yields an
artificially broad or ``defocused'' injection function.
This underlines the importance of
accurate modeling of interplanetary transport effects.  A recent
comparison [\markcite{{\it Earl et al.,} 1995}] showed that nearly
identical results were obtained by three independent computer
codes for treating interplanetary scattering, which used
a Monte Carlo method [\markcite{{\it Earl,} 1987}], a finite
difference method [\markcite{{\it Ruffolo,} 1991}], and an
eigenfunction expansion [\markcite{{\it Pauls and Burger,} 1994}];
the consistency gives one confidence in the numerical accuracy of such
methods.  The transport equation and 
finite difference code of \markcite{{\it Ruffolo} [1995]} also
include convection and adiabatic deceleration in the framework of
focused transport.  Solutions of this transport 
equation by other finite difference codes
[\markcite{{\it Hatzky,} 1996;} \markcite{{\it Lario,} 1997}]
and by a Monte Carlo code (L.\ Kocharov, private communication, 1997)
have corroborated these results.  Again, the agreement makes one 
confident that the various methods are accurately solving the transport
equation.  Remaining systematic errors could arise from the 
underlying transport assumptions.

The goal of the present work is to apply such state-of-the-art
transport simulations and develop semi-au\-to\-ma\-ted fitting
techniques to accurately determine the injection of particles near
the Sun as a function of energy and time.  The fits are 
objective, relying on $\chi^2$ minimization instead of eyeball evaluation
as in much previous work.  Two deconvolution techniques have been tested
for protons from the gradual flare event of July 20, 1981, and the impulsive
flare event of January 2, 1982.  We have successfully fit the data,
and the two techniques yield consistent results.  For the gradual
event of July 20, 1981, the duration of injection was much shorter
for protons of higher energy ($\sim$100 MeV), which we interpret
as indication that the CME shock no longer accelerated protons to
such high energies after traveling beyond a certain distance from the Sun.

\section{2. Numerical Techniques}

To simulate the interplanetary transport of solar energetic
particles, we solve a Fokker-Planck equation of focused
transport that includes the effects of interplanetary scattering,
adiabatic focusing, and solar wind effects, such as solar wind
convection and adiabatic deceleration, to first
order in $(v_{sw}/c)$, where $v_{sw}$ is the (constant)
solar wind speed [\markcite{{\it Ruffolo,} 1995}]:

\begin{flushleft}
\begin{tabular}{lll}
streaming &
$\frac{\stru\pd F}{\stru\pd t}= -\frac{\stru\pd}{\stru\pd z} \mu vF$ 
  \\ [10pt]
convection &
$\mbox{\ \ }-\frac{\stru\pd}{\stru\pd z}\left(1-\mu^2\frac{\stru v^2}
{\stru c^2}\right)v_{sw}\sec\psi F$
  \\ [10pt]
focusing &
$\mbox{\ \ }-\frac{\stru\pd}{\stru\pd\mu}\frac{\stru v}{\stru 2L}
\left[\frac{\stru E'}{\stru E}+\mu\frac{\stru v_{sw}}{\stru v}\sec\psi
\right]$
  \\ [10pt]
&
$\mbox{\ \ \ \ \ }\cdot(1-\mu^2)F$
  \\ [10pt]
differential &
$\mbox{\ \ }+\frac{\stru\pd}{\stru\pd\mu}v_{sw}
\left(\cos\psi\frac{\stru d}{\stru dr}\sec\psi\right)$
  \\ [10pt]
\mbox{\ \ }convection &
$\mbox{\ \ \ \ \ }\cdot\mu(1-\mu^2)F$
  \\ [10pt]
scattering &
$\mbox{\ \ }+\frac{\stru\pd}{\stru\pd\mu}\frac{\stru\varphi}
{\stru2}\frac{\stru\pd}{\stru\pd\mu}\frac{\stru E'}{\stru E}F$
  \\ [10pt]
deceleration &
$\mbox{\ \ }+\frac{\stru\pd}{\stru\pd p}pv_{sw}
\left[\frac{\stru\sec\psi}{\stru2L}(1-\mu^2)\right.$
  \\ [10pt]
&
$\mbox{\ \ \ \ \ }\left.+\cos\psi\frac{\stru d}{\stru dr}(\sec\psi)
\mu^2\right]F,$ 
  \ \ \ \ \ \ \ \ (1) 
\end{tabular}
\end{flushleft}
\setcounter{equation}{1}
\setcounter{table}{0}

\noindent where $F(t,\mu,z,p)\equiv d^3N/(dzd\mu dp)$, the density of 
particles in a given magnetic flux tube (following \markcite{{\it Ng and
Wong} [1979]}) as a function of the four independent variables: $t$ (time),
$\mu$ (pitch angle cosine in the solar wind frame), $z$ (distance along 
the magnetic field), and $p$ (momentum in the solar wind frame).  Also, 
$v$ is the particle speed, $\psi(z)$ is the 
``garden hose angle'' between the magnetic field and the radial direction, 
$L(z)$ is the focusing length, $B/(dB/dz)$, $\varphi(\mu)$ is the pitch
angle diffusion coefficient, and $E'/E = 1-\mu v_{sw}v\sec\psi/c^2$
is the ratio of the total energy in the solar wind frame to that in
the fixed frame.

We assume an Archimedean spiral magnetic field [\markcite{{\it
Parker}, 1958}] for the observed solar wind velocity.  The
pitch angle scattering coefficient, $\varphi$, is parameterized as
$$
\varphi(\mu)=A|\mu|^{q-1}(1-\mu^2),
$$
following \markcite{{\it Jokipii} [1971]}.  We have 
used $q=1.5$, which is in the range of 1.3-1.7 inferred by
\markcite{{\it Bieber et al.}\ [1986]}.
The amplitude, $A$ was determined from $\lambda_\parallel$, the
mean-free path parallel to the field, or $\lambda_r$, the radial mean free
path, which are related by
$$
\frac{\lambda_r}{\cos^2\psi}=\lambda_\parallel=
\frac{3}{(2-q)(4-q)}\frac{v}{A}.
$$
The initial condition places all the particle density at the inner
boundary at $r=0.01$ AU, simulating an instantaneous injection near
the Sun.  Thus the simulation results form a Green's
function for the response to a $\delta$-function injection.
The initial spectrum was estimated from the data.
We use absorbing (zero inflow) boundary conditions at the inner
boundary and at an outer boundary far enough to be inaccessible to
particles during the course of the simulation.

We numerically solved (1) using the computer code of 
\markcite{{\it Ruffolo} [1995]}, as modified to use $t$ instead of $vt$ as 
an independent variable because data gaps can make it awkward to bin the 
data according to $vt$ on a particle-by-particle basis.  
Simulations using $t$ or $vt$ as an independent variable yielded consistent
results.  Each simulation over the time of interest for this work
required $\sim$4 hours on a Sun Ultra-1 workstation.

Since $F$ is defined with respect to $\mu$ and $p$ in the
local solar wind frame, it is necessary to transform $F$ into the
fixed frame to predict counting rates.  This is known as the Compton-Getting
transformation [\markcite{{\it Compton and Getting,} 1935}].  We have fit 
data from the MEH instrument [\markcite{{\it Meyer and Evenson,} 1978};
\markcite{{\it Kroeger,} 1986}] on board the
ISEE 3 spacecraft, which are collected in eight orientational sectors as the
instrument's field of view (half-opening angle of 25$^\circ$) rotates
in the ecliptic plane.  (Because of the narrow field of view of the MEH
instrument, the measurements are essentially restricted to the
ecliptic plane.)  Therefore we used a Monte Carlo simulation to
perform the Compton-Getting transformation and 
calculate a matrix for converting $F(\mu,p)$ into
predicted counting rates in the eight sectors at the energy of interest,
taking into account the geometry of the detector.
The simulation also took into account the predominant magnetic field
direction during the time of interest, rotating the angular
distribution so that $\mu=1$ pointed along that direction.

The simulated and observed count rates have been compared in terms
of the intensity and the anisotropy times intensity.  The intensity is
simply the sum of the eight sectored rates.  We use the anisotropy times 
intensity instead of the anisotropy alone because the product can be
approximated by a linear combination of sectored rates, which is
necessary for one of our deconvolution techniques.  In general,
the observed magnetic field direction and the
axis of symmetry of the particle distribution both fluctuate with
time, and the two do not track each other [\markcite{{\it Bieber
and Evenson,} 1987}].  Presumably the axis of symmetry of the particle
distribution is instead tracking a spatial average of the magnetic
field over the particle gyrations.  This makes it difficult to
precisely predict the direction of the anisotropy, and therefore
we compare the predicted and observed anisotropy times
intensity values that are calculated with respect to the sector with
the highest counting rate.  Setting $\theta=0$ along that sector, a 
first-order harmonic expansion gives 
$F=\langle F\rangle(1+\delta\cos\theta)$, where $\theta$ is the angle
and $\delta$ is the two-dimensional anisotropy in the ecliptic plane.  
Then it can readily be shown that $\delta=2\langle\cos\theta\rangle$, 
where the average is weighted by the particle distribution, and 
the anisotropy times intensity is approximated by 
$2\sum_jF_j\cos\theta_j$, where $F_j$ is the counting
rate in sector $j$.  (If the measurements were evenly distributed in
three dimensions instead of two, it would be approximated by
$3\sum_jF_j\cos\theta_j$.)  Note that the anisotropy times
intensity is calculated in the same manner for both the simulation
results and the observations, so an error in the approximation
should not strongly affect the comparison.

We have developed two techniques for deconvolving the effects of
interplanetary transport in order to determine the underlying time
profile of injection near the Sun, which in turn yields the injected
spectrum as well as the best fit value of the interplanetary
scattering mean-free path.  These techniques solve the inversion problem
\begin{equation}
R(t) = \int_0^t G(t-t')I(t')dt', 
\label{eq:conv}
\end{equation}
where $I(t')$ is the injection of particles versus time
near the Sun, $G(t-t')$ is the Green's function, or the response to a
$\delta$-function injection, which is calculated by the transport
simulation, and $R(t)$ is the ``response,'' i.e., the
measured intensity or anisotropy times intensity at the spacecraft.
Both techniques are semi-au\-to\-ma\-ted in 
that a computer program finds injection parameter values that
minimize the $\chi^2$ of the fit between $R(t)$ and the observational
data (minus background).  Results for different 
values of $\lambda$ are then compared, and the
overall best fit is used as the final result.

The first deconvolution technique finds the optimal piecewise linear
injection function for a given set of ``joints,'' $t_i$.  To illustrate
the technique, we consider fitting the intensity of 27-147 MeV
protons measured by ISEE 3/MEH after the gradual flare event of July
20, 1981 (\callout{Figure 1}a).  The times of the joints in the
piecewise linear injection function, $I(t')$, are chosen {\it a priori}, and
the injection function is constrained to be zero at the first and final 
joints (\callout{Figure 2}c).  Then $I(t')$ is
a linear combination of triangular injections, $I_i(t')$ (Figure 2a).
The first triangular function starts from 0 at the first
joint, rises linearly to 1 at the next joint, and falls linearly to 0
at the following joint (in units of 10$^{26}$ sr$^{-1}$ s$^{-1}$
MeV$^{-1}$, i.e., 10$^{26}$ per unit solid angle of the solar
surface, time, and energy).  The peak time of this function is then the
start time of the next function and so on.  We then
convolve $G(t-t')$, the intensity predicted by the 
transport simulation for an instantaneous injection, with each
$I_i(t')$, which yields the predicted 
response, $R_i(t)$, due to each triangular injection (Figure 2b).  
Now we want to consider the response, $R(t)$, to a
general linear combination of the triangular injections.
Because the transport equation is linear in $F$, the response to a
linear combination of injections, $I(t')=\sum_i a_iI_i(t')$,
is the linear combination of responses, $R(t)=\sum_i a_iR_i(t)$.
Therefore we can use linear least-squares fitting to find the
linear combination that minimizes the $\chi^2$ between $R(t)$
and the observed data (Figure 2d).  Because each
$I_i(t')$ has a peak value of one, the
coefficients, $a_i$, are the values of the injection function at each
joint $t_i$ (Figure 2c); the least squares fit also
directly yields the statistical errors of these values.

Initially, we set $t_0$ to the peak time of the H$\alpha$ flare, set 
$t_1$ and $t_2$ so that $t_2-t_1$ and $t_1-t_0$ were equal
to the width of the time bins of the data, and set further $t_i$ so
that each interval was twice the preceding interval.  The joints, $t_i$, 
were then adjusted according to an objective procedure.  While the procedure
could have been fully automated, we considered it prudent to manually 
execute fits for each set of joints, examining each fit by eye; we never
found it necessary to contradict the decisions mandated by this
procedure.  Each fit runs in the blink of an eye on a Pentium
processor and the fitting procedure is completed in minutes.  

The second deconvolution technique assumes an injection function of the form
$$
I(t) = [C/(t-t_0)]\exp[-A/(t-t_0)-(t-t_0)/B],
$$
where $A$, $B$, $C$, and $t_0$ are free parameters.
This form was originally proposed by \markcite{{\it Reid} [1964]}
as the solution of a coronal diffusion equation over the solar
surface.  However, we adopt this so-called Reid profile only as a convenient 
and widely understood parameterization, and we stress that we do not
assume the existence of coronal diffusion.  

We perform a nonlinear least squares fit in which for each set of 
parameter values, $A$, $B$, and $t_0$,
we numerically evaluate $R(t)$ from (\ref{eq:conv}) using the
results of the transport simulation, and calculate $\chi^2$ for
the fit to the data.  (Given the other parameters, the optimal value
of $C$ can readily be determined.)  The parameters are optimized by
the conjugate direction method [\markcite{{\it Press et al.,} 1988}].
In practice, it was necessary to limit the variation of $A$, $B$, and
$t_0$ to physically reasonable values.
Unfortunately, for nonlinear optimization, one cannot be certain that the 
global minimum has been found, and we had to restart each fit
several times for different initial parameter values. 
Fits were performed for various values of $\lambda$ so as to optimize 
$\chi^2$.  Each fit required $\sim2$ hours on a Pentium processor,
making the entire procedure much slower than the piecewise linear technique.  
Although both deconvolution techniques yielded good fits and consistent 
results, we prefer the piecewise linear deconvolution technique, 
which is faster, does not require any subjective
evaluation by the researcher, and permits a more flexible shape for
the injection function.  

\section{3. Observations and Results}

We have examined data on protons stopping in the ISEE
3/MEH detector.  During the times considered here, ISEE 3 was located
near the inner Sun-Earth Lagrangian point.
Two types of data have been used.  Priority rate (PR)
data consist of raw counts in eight directional sectors over 96 s intervals 
that satisfy the priority 1 logic
(${\rm D1}\times{\rm D3}\times\overline{\rm D2+D13A}\times
\overline{\gamma{\rm H}}\times\overline{\rm D5}\times\overline{\rm D6}$ 
[\markcite{{\it Kroe\-ger,} 1986}]).
It expected that during an intense solar event this rate is
dominated by protons (other ions and electrons have priority 2), 
and detector simulations indicate that the stopping
energies are roughly 27-147 MeV.  Large PR
counting rates may have a very small statistical error, yet
there are still fluctuations that are probably related to magnetic
field irregularities.  In that case, the statistical errors do not reflect
the true level of effectively random rate fluctuations, leading
to large $\chi^2$ values even for a reasonable fit.
Pulse height (PH) data enable one to reject background events more cleanly 
and also to determine the energy (within $\sim1$ MeV) and time of arrival 
of each particle.  While such data are clearly more desirable, the 
transmission of PH data to Earth was constrained by telemetry limits and 
depended on the priority logic.  One must correct for this, and during
times with a large electron flux, the PH data may yield poor statistics
for the lower-priority protons.

We chose to test the deconvolution techniques using data for two
solar events, one impulsive (X ray duration $<1$ hour) and one gradual
(X ray duration $>1$ hour).  Thus the objective is to examine events for 
which the particles are evidently of solar origin, and are sufficiently 
intense to permit a detailed analysis and for which there is an evident 
anisotropy, which is an aspect of the fitting we would like to test.
Other considerations are the absence of data gaps at the very start
of the event and a small coronal distance between flare site and the
footpoint of the (average) magnetic field connected to the
spacecraft.  We chose to examine the gradual event of July 20, 1981, and
the impulsive event of January 2, 1982.  Starting with such 
``well-connected'' events simplifies the interpretation of the results; in 
future applications one could use these deconvolution techniques for 
poorly connected events in order to study lateral transport mechanisms.  

Figure 1 shows PR proton data for the gradual event of July 20, 1981.
This event was associated with an H$\alpha$ flare at 25$^\circ$S, 
75$^\circ$W, peaking at 1322-1336 UT (based on two observatories, 
Solar-Geophysical Data).  The X ray decay time was 67 min 
[\markcite{{\it Cliver et al.,} 1989}], indicating a gradual flare.  
The solar wind speed was approximately 375 km s$^{-1}$ 
(S.\ Bame, private communication via ISEE 3 data pool, 1981).
The magnetic field varied rapidly in both magnitude and
direction during the time of interest 
(E.\ Smith, private communication via ISEE 3 data pool, 1981).

Although the intensity was rather smooth as a function of time,
the anisotropy times intensity suddenly fell by a factor of about
4 at 1600 UT and recovered at 1700 UT.  Such sudden disappearances of
anisotropy have been observed for other flare events by
\markcite{{\it Evenson et al.}\ [1982]}.  As discussed earlier, the
anisotropy vector closely follows the magnetic field direction, so it
would be expected to be especially sensitive to erratic magnetic
field fluctuations.  In fact, the field magnitude dropped particularly
sharply during 1600-1700 UT, corresponding to a very short focusing
length, $L\equiv -B/(dB/dz)\approx -0.04$ AU; this strong reverse
focusing apparently negated most of the outward-going anisotropy during
that time.  It is impossible for a transport model based on an
Archimedean spiral field to predict such drastic fluctuations in the 
anisotropy.  Therefore we conclude that when the magnetic field is erratic,
as for this flare, our transport simulations are only appropriate for 
fitting the intensity as a function of time and particle energy.

For the proton event of July 20, 1981, there were sufficient statistics to 
use PH data and to separately examine different proton energy ranges.
Fits were performed using both a piecewise linear injection profile and a
Reid profile, assuming a position-independent radial mean free path, 
$\lambda_r$, as recommended by \markcite{{\it Palmer} [1982]}.
The resulting fit parameters are given in \callout{Tables 1 and 2}, and 
piecewise linear injection functions are shown in \callout{Figure 3}.  
In each case, both deconvolution techniques yielded the same best fit value 
of $\lambda_r$, except for the highest energy bin where they differed
slightly.  There is a hint of an increasing trend of $\lambda_r$ with
energy, which is consistent with previous results and theoretical
calculations [e.g., \markcite{{\it Dr\"oge et al.,} 1997};
\markcite{{\it Schmidt and Dr\"oge,} 1997}]
The injection functions can be compared in terms of the 
full width at half maximum (FWHM).  While the Reid profile
always has a higher FWHM, the energy dependence is similar for both methods.  

Both deconvolution techniques indicate a much narrower injection
profile for the highest energy bin (75-147 MeV).  As discussed
earlier, there is evidence that for gradual events the acceleration
of ions takes place at a CME-driven shock as it propagates outward
through the corona.  Therefore the injection as a function of time
can also be interpreted as injection as a function of distance
[\markcite{{\it Kahler et al.,} 1990}], though the CME speed is not
known for this particular event.  Thus these results suggest that for 
this event the CME/shock system lost the ability to accelerate
particles to $\sim$100 MeV after traveling beyond a certain distance
from the Sun.  

By integrating the injection function over time for each energy bin, 
we can estimate the spectrum of emitted particles.  We compare our
techniques with the commonly used time of maximum (TOM) method, in which one
uses the observed peak intensity for each energy interval to estimate
the relative spectrum.  For the July 20, 1981, event, we find that the
piecewise linear and Reid injection profiles yield very similar
absolute spectra, except at the highest energies (\callout{Table 3}).  
The relative spectra of both methods are also similar to the TOM spectrum, 
except at the high energy bin.  The TOM technique implicitly assumes a
similar injection profile and scattering mean-free path for each
energy, so one might well expect some deviation for the high energy
bin, where both deconvolution techniques yield a much shorter
duration of injection.  Also, there were large statistical errors for
this energy bin.  From our results we conclude that the TOM
relative spectrum can be accurate for a short energy span (here a
factor of 3), but for wider energy ranges it is worthwhile to 
determine the spectrum more accurately by deriving the
energy-dependent injection profiles.  Note that the TOM method does not 
yield an absolute spectrum, and the consistency of our two methods for the 
absolute spectrum improves our confidence in both of them.

We have also analyzed PR data from the impulsive flare event of
January 2, 1982 (\callout{Figure 4}).  This event was associated with an 
H$\alpha$ flare at 18$^\circ$S, 88$^\circ$W, peaking at 0620-0621 UT 
(Solar-Geophysical Data).  The X ray decay time was 16 min 
[\markcite{{\it Cliver et al.,} 1989}], and the solar wind speed was 
approximately 350 km s$^{-1}$ (S.\ Bame, private communication via ISEE 3 
data pool, 1982).  The magnetic field was stable in magnitude and
direction during the time of interest 
(E.\ Smith, private communication via ISEE 3 data pool, 1982).
Because of this, we were able to analyze the anisotropy in addition
to the intensity for this event.  Unfortunately, because of the limited
statistics and strong intensity of electrons (which had a higher
priority for pulse-height telemetry), we were unable to examine PH data 
in detail to determine energy-dependent injection functions.

Simultaneously fitting the intensity and anisotropy times intensity 
provides a stringent test of the injection and transport models.
Our two techniques for deconvolving the effects of interplanetary
transport yielded similar FWHM durations of injection (17 and 24
min, respectively; see Tables 1 and 2) and yielded reasonable
fits to the intensity and anisotropy times intensity.
The duration of injection over this broad energy range (27-147 MeV)
was similar to the decay time of X ray emission (16 min) and was 
markedly shorter than the corresponding duration for the gradual
event of July 20, 1981, though that event had a similarly short
injection duration for the highest energies.

We close this section with examples of how less accurate transport 
assumptions can artificially broaden the derived injection profile.  
For an analogy, consider the deconvolution of a telescope image to
account for the point spread function.  An inaccurate estimate of the
point spread function yields a deconvolved image that is still
artificially broad compared to the true image size.  When an improved
point spread function is used, one obtains a sharper image.

\callout{Figure 5} shows examples of a similar effect for our
deconvolution problem, in which we compare fits to PR data from both
events, assuming that either $\lambda_r$ or $\lambda_\parallel$ is
constant in position.  Previous authors [\markcite{e.g., {\it
Palmer,} 1982}] have concluded that a constant $\lambda_r$ is a better and
reasonable assumption for transport in the inner heliosphere.
We find that the fits assuming a constant $\lambda_\parallel$ (Figures
5c and 5d) yield much broader injection functions than those
based on the presumably more accurate assumption of a constant $\lambda_r$
(Figures 5a and 5b).  These examples stress the importance of
using an accurate transport model when determining the injection
function near the Sun.

\section{4. Discussion}

The previous study with goals most similar to ours was that of
\markcite{{\it Ma Sung and Earl} [1978]}, which employed approximate
analytic solutions to a focused transport equation.  One of their
assumptions was a position-independent ratio, $\lambda_\parallel/L$,
where $L$ is the focusing length.  This assumption was necessary for their
analytic approximation but is less
accurate than the assumption of a constant $\lambda_r$ and a function
$L(z)$ derived for an Archimedean field.
Based on our results, one would expect injection
functions that are smeared out when compared with those for a
constant $\lambda_r$ (Figure 5).  Those authors frequently found
particle release times of the order of hours, and it is possible that
their injection profiles were artificially broad due to that
assumption.  The deconvolution techniques presented here, along with
more realistic transport assumptions and modern numerical simulations, 
could be profitably applied to survey the injection profiles of a
variety of events, as \markcite{{\it Ma Sung and Earl} [1978]} did.
Shortly, we will consider how such results should be
interpreted in light of our modern understanding of the origin of
solar energetic particles.

A valid question is how one knows that injection occurs
near the Sun, when the interplanetary shocks that frequently accompany
gradual events are known to be capable of accelerating particles 
[\markcite{{\it Gosling et al.,} 1981}].  For the gradual event considered 
here, several days after the flare, the log(proton intensity) versus time 
showed a bump (July 23) and then a double-peaked increase (July 24-25) to 
fluxes $<$1/30 of the initial peak.  There was intense geomagnetic activity 
at these times (Solar-Geophysical Data).  
Presumably these features are associated with the passage of a CME/shock
system, which affected the particle propagation or accelerated particles to
higher energies.  In this work, we focus on the proton observations
on July 20-21, 1981, well before such features arose.  The rapid rise
and exponential decline are consistent with injection shortly after
the flare occurrence convolved with the effects of interplanetary
transport.  Since the CME could not have traveled far from
the Sun during that time, we concluded that the emission was near the
Sun and neglected any source motion.  

We note further that all measurements of ionic charge states above 3 
MeV/nucleon indicate that solar energetic ions are accelerated out of 
coronal material [\markcite{{\it Boberg et al.,} 1996}].  
This implies that the CME/shock system first accelerates
ions out of coronal material, presumably while it is still near the
Sun, and can also further accelerate ions as it propagates outward.
The relative importance of acceleration at different distances from
the Sun undoubtedly varies from event to event and varies with energy.  

Our results indicate that for the gradual event of July 20, 1981, the
duration of injection was much shorter for higher energies (Figure 3), 
which implies that the CME/shock system lost the ability to
accelerate a significant flux of $\sim$100 MeV protons after
traveling a certain distance from the Sun.  
We also note that the shock-associated particle increases on July 23,
and July 24-25 had a significantly steeper spectrum than the main
peak on July 20 and were not seen at all in the high energy bin.  
Therefore one might interpret the short duration
of injection at higher energy as indicating a transition from
injection with a harder spectrum (when the CME shock is close to the Sun)
to injection with a softer spectrum (as for the interplanetary shock).
While it is imprudent to draw
general conclusions from one flare/CME event, confirmation of this
result for other events would provide important information on the
acceleration mechanism for escaping energetic ions from such events.

It is also interesting to compare the best fit values of $\lambda$
for fits assuming a constant $\lambda_\parallel$ or a constant
$\lambda_r$, which are largely determined by the observed intensity
decay at long times.  We obtain ratios of $\lambda_\parallel/\lambda_r=5.8$
and 4.5 (Table 1).  At any given point, 
$\lambda_\parallel/\lambda_r=\sec^2\psi$, so the observed ratios are 
characteristic of $r=1.9$ and 1.5 AU.  This indicates that the intensity
versus time of solar particles at 1 AU is strongly influenced by the
transport conditions at greater distances.

Since we have only treated the interplanetary transport parallel to the 
magnetic field, we have derived injection profiles under the assumption that 
particles are strictly confined to a narrow flux tube connecting the 
observer to the source near the Sun.  
However, perpendicular diffusion implies that the observed particles
could have originated from a distribution of longitudes and latitudes.
Summarizing a variety of results, \markcite{{\it Palmer} [1982]}
recommends using $\kappa_\perp=(v/c)\times 10^{21}$ cm$^2$ s$^{-1}$ at
1 AU (well away from sector boundaries, corotating interaction regions,
etc.).  For our typical proton velocity of $c/3$, 
this implies an angular spread of $\sim$7$^\circ$ after 3 hour or
$\sim$20$^\circ$ after 1 day.
In comparison, gradient and curvature drifts imply an angular motion
$\sim$0.1 AU day$^{-1}$ near the Earth or an integrated
angular motion $\sim$0.1$^\circ$ for the first particles that
arrive and are hence negligible during the time of interest, as is
the rotation of the Sun.

For gradual events, there is evidence that coronal mass ejection
shocks can accelerate and release particles over a large fraction of the 
solar surface [\markcite{{\it Mason et al.,} 1984;} \markcite{{\it
Kahler,} 1992}], so our results are really telling us the injection
profile averaged over $\sim 7^\circ$ of solar latitude and
longitude from the footpoint 
magnetically connected to the observer (which is itself uncertain,
except when there are observations of moving interplanetary 
type III radio bursts that track the mean motion of electrons along
the interplanetary magnetic field [\markcite{{\it Reames and Stone,}
1986;} \markcite{{\it Reiner et al.,} 1995}]).  The accuracy of the
measured duration of injection as a function of energy 
should be unaffected insofar as perpendicular
diffusion does not significantly increase the distance traveled
before particles actually reach the observer, i.e., as long as it is
much weaker than parallel diffusion.

For impulsive events,
the interpretation is potentially more complex: Since particles are
accelerated at the flare site, their lateral spread could be due to
perpendicular diffusion in the interplanetary magnetic field
(including the ``random walk'' of the magnetic field lines themselves 
[e.g., \markcite{{\it Jokipii,} 1966;} \markcite{{\it Matthaeus et al.,}
1995}]) or to lateral transport within the corona.  For impulsive
flares that are magnetically well-connected to the observer, such as
the event of January 2, 1982, which we have considered here, we can
again argue that our determination of the duration of injection is
accurate as long as perpendicular diffusion does not
significantly increase the propagation time.  However, the absolute
normalization of the observed flux depends strongly on the lateral
extent of the particle distribution at the observer's radius.  In
this case, we are determining the number of particles injected per solid
angle at the observer's radius.

\section{5. Conclusions}

We have developed two techniques for deconvolving the effects of
interplanetary transport using numerical solutions of the
transport equation: (1) assuming a piecewise linear injection
profile and (2) using a Reid injection profile.  The deconvolution can yield
the interplanetary scattering mean free path, the
injection profile as a function of energy, and the spectrum.
The two techniques yield consistent results for the gradual flare/CME
event of July 20, 1981, and the impulsive flare event of January 2,
1982, giving us confidence in both techniques.
The piecewise linear profile is preferred because the
fitting procedure is faster, objective (not relying on user
evaluation of fits), and permits a more flexible profile shape.
It is important to examine the observed magnetic field; if this 
is erratic, it may not be possible to fit the anisotropy data with
simple transport assumptions.  For the July 20, 1981, event, a simple 
time-of-maximum estimate of the spectrum agreed with our
results over a factor of 3 in energy but deviated at higher energies.
The FWHM of injection was much shorter for higher
energies ($\sim$100 MeV), indicating that for this event, the CME and 
associated shock lost their efficiency of accelerating such particles after
traveling a certain distance from the Sun.

\vfill\eject

\acknowledgments
The authors would like to thank Paul Even\-son and Peter Meyer for 
stimulating discussions and for providing the ISEE 3 data.  We are also 
grateful for useful discussions with Wolfgang Dr\"oge and John Bieber.
This work was partially supported by grants from Chulalongkorn
University's Rachadapisek Sompoj Fund and the Thailand Research Fund.
TK also received support from a Naresuan University Scholarship.

The Editor thanks M.\ Aschwanden and another referee for their
assistance in evaluating this papaer.

\end{article}

\begin{figure}
\caption{
Intensity and anisotropy times intensity as a function of time for
27-147 MeV protons following the July 20, 1981, gradual solar flare
event.
}
\end{figure}

\begin{figure}
\caption{Illustration of the deconvolution technique
for a piecewise linear injection function near the Sun.
The transport equation is solved for an instantaneous injection
of particles.  The resulting Green's function is convolved with 
(a) triangular injection profiles to (b) yield response functions.
Linear, least squares fitting yields (d) the linear combination 
of response functions that best fits the data 
and (c) the corresponding best-fit piecewise linear injection profile.  
}
\end{figure}

\begin{figure}
\caption{
Fits to the (a) observed proton intensity versus time 
in four energy bins for (b) optimal piecewise linear 
injection profiles.  Note the expanded timescale in the right panels.
}
\end{figure}

\begin{figure}
\caption{
Fits to the (a) observed 27-147 MeV proton intensity versus time 
and anisotropy times intensity versus time
for the (b) optimal piecewise linear injection profile.  
Note the expanded timescale in Figure 4b.
}
\end{figure}

\begin{figure}
\caption{
Comparison of best-fit piecewise linear injection profiles for
transport simulations assuming either (a-b) $\lambda_r$ or
(c-d) $\lambda_\parallel$ to be independent of position
for 27-147 MeV protons on July 20, 1981, for Figures 5a and 5c
and January 2, 1982, for Figures 5b and 5d.
The assumption of a constant $\lambda_r$ is expected to be more
accurate, and it yields a sharper injection function for both events.
}
\end{figure}

\clearpage

\begin{table}
\caption{Fit Parameters for a Piecewise Linear Injection Profile}
\vspace{5pt}
\begin{tabular}{lcccclrcc}
\tableline
& & & & & & & & \\[-6pt]
Date & Data & Energy, & $\lambda_r$ or $\lambda_\parallel$ & 
  $\lambda,$ & \multicolumn{1}{c}{$\{a_i\}$\tablenotemark{a}} 
\tablenotetext{a}{Units of 10$^{26}$ sr$^{-1}$ s$^{-1}$ MeV$^{-1}$.}
  & \multicolumn{1}{c}{$\chi^2$/d.f.} & FWHM, & Figure \\
& & \multicolumn{1}{c}{MeV} & constant? & AU & & &
  min & \\
& & & & & & & & \\[-6pt]
\tableline
& & & & & & & & \\[-6pt]
July 20, 1981 & PH & 27-39 & $\lambda_r$ & 
  0.08 & \{7.2,\,0.5,\,0.7\} & 1.92 & \phantom{0}33 & 3 \\
& PH & 39-55 & $\lambda_r$ & 
  0.10 & \{1.6,\,1.9\}     & 0.96 & \phantom{0}65 & 3 \\ 
& PH & 55-75 & $\lambda_r$ & 
  0.12 & \{0.57,\,0.93\}   & 1.11 & \phantom{0}52 & 3 \\ 
& PH & 75-147 & $\lambda_r$ & 
  0.10 & \{0.58,\,0.04\}   & 1.24 & \phantom{0}12 & 3 \\ 
& PR & 27-147 & $\lambda_r$ &
  0.12 & \{1.095,\,0.089,\,0.120\} & 161.65 & \phantom{0}31 & 2, 5 \\
& PR & 27-147 & $\lambda_\parallel$ &
  0.70 & \{0.296,\,0.503,\,0.414,\,0.0942\} & 89.68 & 156 & 5 \\
Jan.\ 2, 1982 & PR & 27-147 & $\lambda_r$ &
  0.20 & \{12.0,\,2.7\}$\times10^{-3}$ & 2.09 & \phantom{0}17 & 4, 5 \\
& PR & 27-147 & $\lambda_\parallel$ &
  0.90 & \{2.8,\,4.1,\,3.3,\,2.1,\,1.0\}$\times10^{-3}$ & 
  2.35 & \phantom{0}52 & 5 \\
& & & & & & & & \\[-6pt]
\tableline
\end{tabular}
\end{table}


\begin{table}
\caption{Fit Parameters for a Reid Injection Profile}
\vspace{5pt}
\begin{tabular}{lcccccrccc}
\tableline
& & & & & & & & & \\[-6pt]
Date & Data & Energy, & $\lambda_r,$ & $A,$ & $B,$ &
  \multicolumn{1}{c}{$C$\tablenotemark{a}} & $t_0,$ &
  $\chi^2$/d.f. & FWHM, \\
& & \multicolumn{1}{c}{MeV} & AU & min & min & &
  UT & & min \\
& & & & & & & & \\[-6pt]
\tableline
& & & & & & & & \\[-6pt]
July 20, 1981 & PH & 27-39 &  
  0.08 & 23.6 & 81.7 & 436.8 & 1330 & 2.43 & 47 \\
& PH & 39-55 &  
  0.10 & 96.0 & 44.5 & 1586.4 & 1314 & 1.11 & 68 \\
& PH & 55-75 &  
  0.12 & 54.6 & 46.7 & 251.0 & 1325 & 1.18 & 55 \\
& PH & 75-147 &  
  0.12 & 22.6 & 30.0 & 30.5 & 1324 & 0.99 & 29 \\
Jan.\ 2, 1982 & PR & 27-147 & 
  0.20 & 96.7 & \phantom{0}9.0 & 165.4 & 0535 & 3.70 & 24 \\
& & & & & & & & 
  \tablenotetext{a}{Units of 10$^{26}$ sr$^{-1}$ s$^{-1}$ MeV$^{-1}$.}
  \\[-6pt]
\tableline
\end{tabular}
\end{table}

\begin{table}
\caption{Spectra of Protons Injected on July 20, 1981}
\vspace{5pt}
\begin{tabular}{lcccc}
\tableline
& & & & \\[-6pt]
Technique & \multicolumn{4}{c}{Energy Range, MeV}\\[.3ex]
\cline{2-5}\\[-1.6ex]
& 27-39 & 39-55 & 55-75 & 75-147 \\
& & & & \\[-6pt]
\tableline
& & & & \\[-6pt]
\multicolumn{5}{c}{{\it Absolute Spectra,} 
                   10$^{26}$ sr$^{-1}$ s$^{-1}$ MeV$^{-1}$}\\
& & & & \\[-6pt]
Piecewise linear & 19,000 & 7120 & 2770 & 465\\
Reid profile     & 19,900 & 7120 & 2810 & 581\\
& & & & \\[-6pt]
\multicolumn{5}{c}{\it Relative Spectra}\\
& & & & \\[-6pt]
Piecewise linear & 1 & 0.374 & 0.145 & 0.0244\\
Reid profile     & 1 & 0.358 & 0.141 & 0.0292\\
Time of maximum  & 1 & 0.343 & 0.136 & 0.0185\\
& & & & \\[-6pt]
\tableline
\end{tabular}
\end{table}

\clearpage

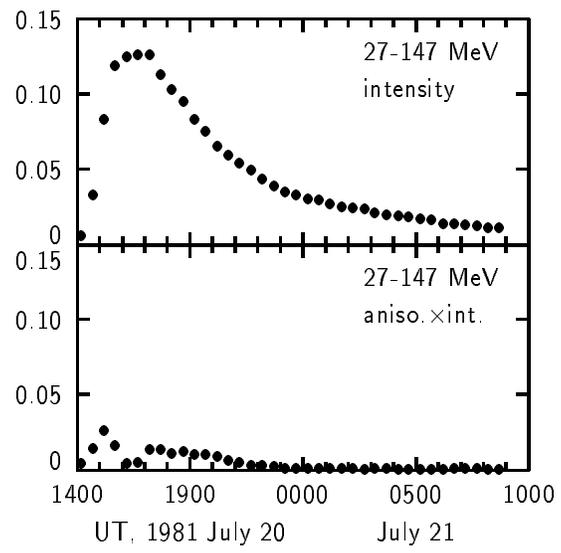
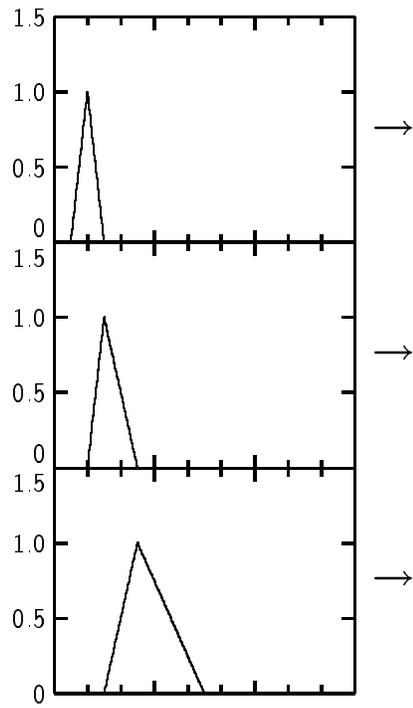
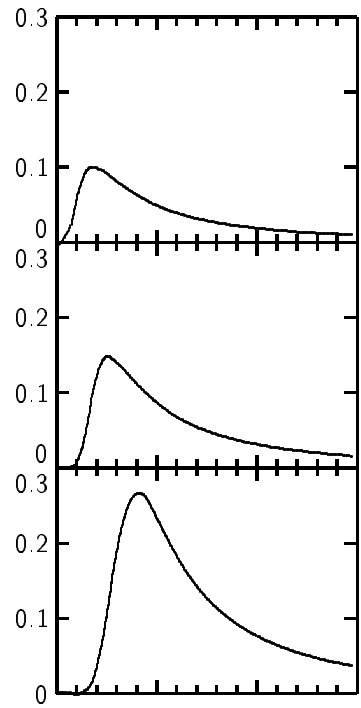
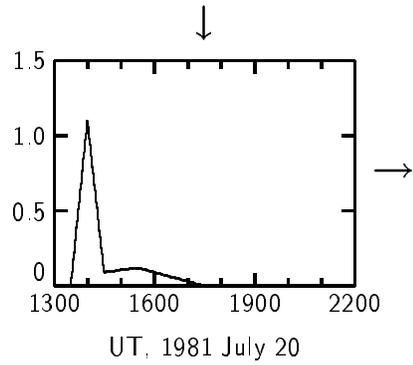
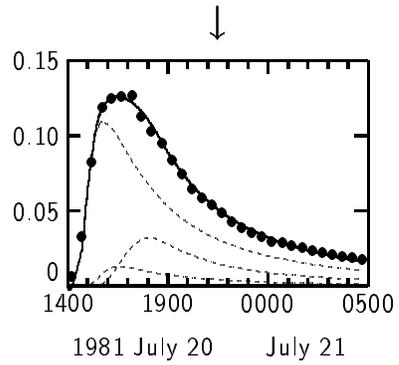
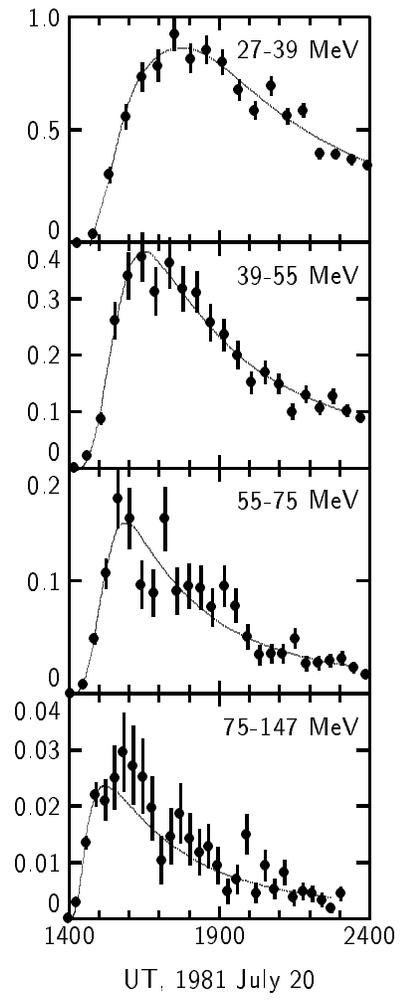
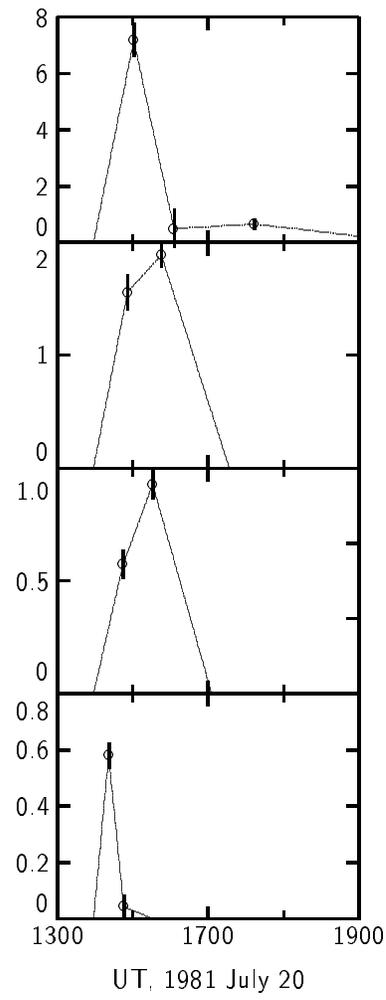
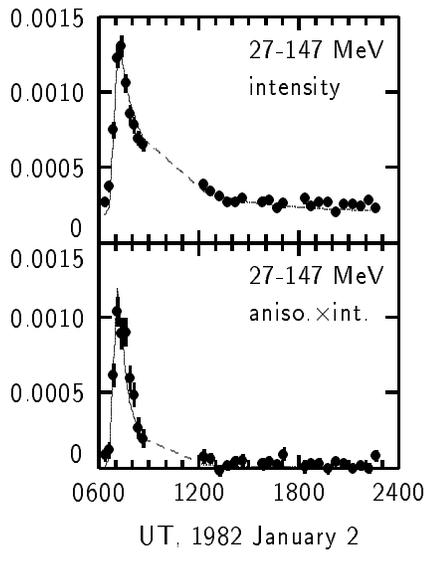
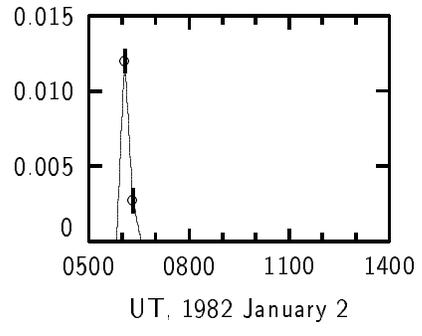
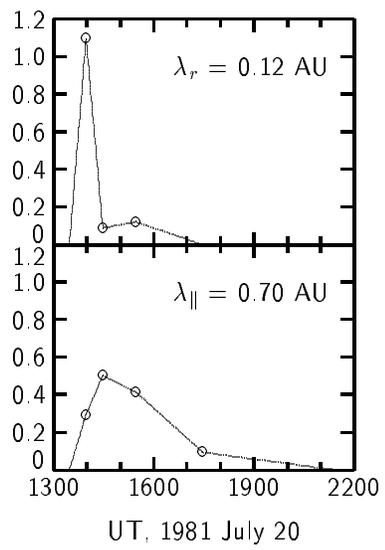
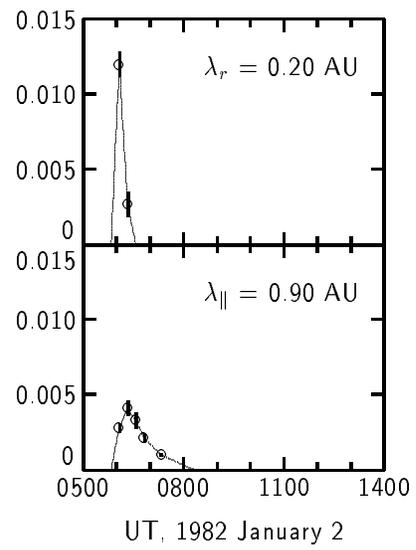
\end{document}